# Low Temperature Spin Freezing in $Dy_2Ti_2O_7$ Spin Ice

J. Snyder[1], B. G. Ueland[1], J. S. Slusky[2], H. Karunadasa[2], R. J. Cava[2], and P. Schiffer[1*]

[1]*Department of Physics and Materials Research Institute, Pennsylvania State University, University Park PA 16802*

[2]*Department of Chemistry and Princeton Materials Institute, Princeton University, Princeton, NJ 08540*

**Abstract**

We report a study of the low temperature bulk magnetic properties of the spin ice compound $Dy_2Ti_2O_7$ with particular attention to the ($T < 4$ K) spin freezing transition. While this transition is superficially similar to that in a spin glass, there are important qualitative differences from spin glass behavior: the freezing temperature increases slightly with applied magnetic field, and the distribution of spin relaxation times remains extremely narrow down to the lowest temperatures. Furthermore, the characteristic spin relaxation time increases faster than exponentially down to the lowest temperatures studied. These results indicate that spin-freezing in spin ice materials represents a novel form of magnetic glassiness associated with the unusual nature of geometrical frustration in these materials.

[*]*schiffer@phys.psu.edu*

## I. Introduction

Geometrically frustrated magnetic materials, in which the topology of the spin lattice leads to frustration of the spin-spin interactions, have been demonstrated to display numerous unusual cooperative spin states [1,2]. Of particular recent interest are the rare earth pyrochlores such as $Dy_2Ti_2O_7$, $Ho_2Ti_2O_7$, and $Ho_2Sn_2O_7$ [3,4,5,6,7,8,9,10,11,12,13,14,15,16], in which the lattice geometry and spin symmetry lead to frustration of ferromagnetic and dipolar interactions [17,18,19,20]. The magnetic rare-earth ions in these materials are situated on a lattice of corner-sharing tetrahedra, where their spins are constrained by crystal field interactions to point either directly toward or directly away from the centers of the tetrahedra. To minimize the dipole and ferromagnetic exchange interactions, the spins on each tetrahedron must be oriented such that two spins point inward and two point outward in exact analogy to the constraints on the positions of the hydrogen atoms in the ground state of ice [21,22,23]. The resulting high degeneracy of spin states leads to a disordered low temperature state analogous to that of ice, and the low temperature magnetic state in these materials has thus been termed "spin ice".

The spin ice state has been demonstrated experimentally through neutron scattering studies [5,13,24] and also through measurements of the magnetic specific heat [7,15]. The latter yield a measured ground state spin entropy in good agreement with the theoretical prediction for the "ice rules" (first codified by Pauling) and experimental results for ice [7,15,23]. While the spin entropy only freezes out below $T_{ice}$ ~ 4 K in $Dy_2Ti_2O_7$, a.c. magnetic susceptibility studies show a strongly frequency dependent spin-freezing at $T$ ~ 16 K [8,10], below which the high frequency susceptibility ($f > 100$ Hz) is suppressed. Because of the high degree of structural and chemical order in this material, spin relaxation in the vicinity of the $T \sim 16$ K spin-freezing is associated with a very narrow distribution of spin relaxation times (determined from the frequency dependence of the imaginary part of the a.c. magnetic susceptibility). This distribution is sufficiently narrow that the spin relaxation can be characterized by a single temperature-dependent relaxation time $\tau(T)$ which is thermally activated for $T > T_{cross}$ (where $T_{cross} \sim 13$ K) and exhibits relatively weak temperature dependence for $T_{cross} > T > T_{ice}$ -- due to a crossover from thermal to quantum spin relaxation at $T \sim T_{cross}$ [9,25]. The weak temperature dependence of $\tau(T)$ for $T_{cross} > T > T_{ice}$, is responsible for the absence of freezing in low frequency susceptibility in that temperature range [10,25].

For $T < T_{ice}$, a.c. susceptibility measurements show a second frequency dependent spin-freezing [8,10,11,26] which corresponds to the loss of entropy observed in the specific heat measurements [7,15]. This freezing is more complete than that at higher temperature, i.e. the a.c. susceptibility goes to zero below the frequency dependent freezing temperature ($T_f$) for all measured frequencies, and there is an associated bifurcation between the field-cooled and zero-field cooled magnetization [10,11,13]. This lower temperature freezing is thus directly analogous to freezing into a spin glass state in disordered frustrated magnets. While experimental studies have examined the local magnetic structure within the low temperature frozen state [4,5,15] as well as field-induced transitions to a polarized state [14], there has been no direct comparison of the bulk magnetic properties near the low temperature spin-ice freezing to those associated with spin-glass freezing encountered in disordered frustrated magnets [27]. We report a detailed study of the bulk magnetic properties of $Dy_2Ti_2O_7$ in the low temperature regime of $T < T_{ice}$, and we find that they differ significantly from those of a spin glass.

## II. Experimental details

Polycrystalline $Dy_2Ti_2O_7$ samples were prepared using standard solid-state synthesis techniques described previously [8,28]. X-ray diffraction demonstrated the samples to be single-phase, and Curie-Weiss fits done to the high temperature susceptibility were consistent with $J = 15/2$ $Dy^{3+}$ ions. We study the magnetization ($M$) as well as the real and imaginary parts ($\chi'$ and $\chi''$) of the a.c. susceptibility ($\chi_{ac}$). We measured $M$ with a Quantum Design MPMS SQUID magnetometer for temperatures above 1.8 K. At lower temperatures, we measured magnetization with a capacitive field-gradient magnetometer assembled from a sapphire base and a quartz paddle (Ferro-Ceramic Grinding Inc.) and mounted in a sample can filled with superfluid $^4$He for thermal contact [28]. For temperatures above 1.8 K, we measured $\chi_{ac}$ with the ACMS option of the Quantum Design PPMS cryostat, while at lower temperatures we used a simple inductance coil in a dilution refrigerator [28]. All samples studied were potted in non-magnetic epoxy (Stycast 1266). This allowed for reliable thermal contact in the case of the a.c. susceptibility studies and control over the shape of the sample in the case of the magnetization measurements. The potting in epoxy had only minor effects on the measured properties, as demonstrated below and discussed in detail elsewhere [28].

## III. Experimental results

### A. Magnetization

The $T > 1.8$ K magnetization indicated a small ferromagnetic Weiss constant of $\Theta_w \approx 0.2$ K, which is quite close to the value of $\Theta_w \approx 0.5$ K reported by Ramirez *et al*. [7]. We measured the low temperature magnetization on warming from 100 mK to 1.2 K at a rate of ~ 5 mK/min. Zero field cooled (ZFC) and field cooled (FC) data are shown in figure 1. A bifurcation was seen between the FC and ZFC magnetization at $T$ ~ 650 mK. Below this temperature, the FC magnetization is completely reversible in temperature at a fixed field while the ZFC magnetization is irreversible. This signifies the onset of spin freezing on the time scale of the magnetization measurement (~ $10^2$ - $10^3$ seconds). One of the closely studied properties of conventional spin glasses is how the freezing temperature (commonly taken as the bifurcation point between the FC and ZFC magnetization) evolves with applied magnetic field. In conventional spin glasses, a sufficiently strong applied field quenches the glass state, and the temperature at which the glass state appears decreases monotonically and usually quite rapidly with increasing applied magnetic field (the so-called AT or GT lines [27]). As shown in figure 1, the temperature of this bifurcation in $Dy_2Ti_2O_7$ is only weakly dependent on applied field (the application of a magnetic field actually slightly *increases* the freezing temperature observed in the a.c. susceptibility data discussed below). On the other hand, the percentage difference between the two data sets decreases with applied field, and above 5 kOe we observed no difference between the FC and ZFC magnetization.

The spin ice freezing is also reflected in hysteresis in the field dependence of the magnetization below 650 mK. Before cycling the field, the sample was zero-field-cooled from 1.2 K to the temperature of the measurement. The magnetization $M(H)$ was then measured as the field was swept up to 10 kOe, down to –10 kOe, and back up to 10 kOe to close the loop. As can be seen in figure 2, the loop is almost identical at 250 mK and 400 mK, with a width of ~ 4 kOe. As expected, $M(H)$ becomes reversible again above 650 mK (as evidenced by the loop being closed completely at 800 mK).

The irreversibility in the low-temperature spin state was further studied by examining the remanent magnetization in zero field. The thermoremanent magnetization (TRM) was found by cooling the sample from 1.2 K in a field to the desired temperature, reducing the field to zero at a rate of 0.1 T / min, and then measuring the magnetization as a function of time to obtain the asymptotic moment. The isothermal remanent magnetization (IRM) was found by cooling the sample from 1.2 K in the absence of a field, and then cycling the field from $0 \rightarrow H \rightarrow 0$ and measuring the magnetization as a function of time to obtain the asymptotic moment. The sample was held at H for at least 8 hours to obtain nearly complete relaxation in the field. Our measurements of the IRM and TRM as a function of applied field can be seen in figure 3, and are qualitatively consistent with expectations for a spin glass [27]. The difference between the IRM and TRM below 5 kOe represents the fact that the system retains a "memory" of its preparation even under the same final conditions. The saturation of IRM and TRM above 5 kOe shows that sufficiently high fields can destroy this "memory", consistent with the equivalence of the ZFC and FC $M(T)$ data taken in fields above 5 kOe.

**B. A.C. Magnetic Susceptibility**

In contrast to magnetization studies, a.c. susceptibility measurements with varying frequency allow a direct probe of the spin relaxation time. The characteristic behavior of the high temperature a.c. susceptibility is shown in figure 4, in which the freezing at $T \sim$ 16 K is evident as well as the maximum in $\chi'(T)$ at $T < 4$ K which is associated with the development of correlations for $T < T_{ice}$ (manifested in the irreversibility of the magnetization below 650 mK described above). These data also demonstrate the relatively small difference between loose powder samples and the samples potted in epoxy being studied here. In order to examine the spin relaxation process in detail a.c. susceptibility data were taken in the vicinity of this low temperature freezing over a relatively broad range of low frequencies ( $0.1 < f < 500$ Hz).

With the magnetization showing a bifurcation at $T \sim 650$ mK, we expect the a.c. susceptibility in our frequency range to freeze out above this temperature (as has been observed previously in a more limited frequency range [8,10,11]). As shown in figure 5, $\chi'(T)$ does have a maximum and then drops to zero below 0.7 K for all measured frequencies. We also observe the rise in $\chi''(T)$ corresponding to the maximum in $\chi'(T)$, as expected from the Kramers-Kronig relations.

The $T < T_{ice}$ spin freezing feature in $\chi'(T)$ is quite broad, as was also the case for the freezing observed near $T = 16$ K. We define a freezing temperature, $T_f$, for the lower temperature feature as the maximum in $\chi'(T)$. The data taken in the dilution refrigerator with those taken in the PPMS are combined in Figure 6 to examine the dependence of $T_f$ on frequency and magnetic field. The a.c. data taken in a magnetic field show that a field enhances $T_f$, which is consistent with the behavior seen at the higher temperature spin-freezing [8]. Unlike the higher temperature feature, a reasonable extrapolation of $T_f$ to very long times does approach the bifurcation temperature seen in magnetization measurements. However, the frequency dependence of $T_f$ cannot be fit to an Arrhenius law ($f = f_0 e^{-E_a/k_B T_f}$), suggesting that this relaxation is not simply thermally activated. Such non-Arrhenius behavior has previously been observed in the dilute Ising spin system LiHo$_{1-x}$Y$_x$F$_4$ [29], and is consistent with the previously suggested importance of quantum spin relaxation in this system [9,25],

To further characterize spin relaxation time in Dy$_2$Ti$_2$O$_7$, we also measured $\chi''(f)$ at temperatures from 0.8 K to 1.8 K, as shown in figure 7. Like the higher temperature data, $\chi''(f)$ displays a single, relatively sharp peak which implies that there is a narrow range of relaxation times or effectively a single characteristic relaxation time, $\tau$, for the spins in zero field (where $1/\tau$ is the frequency of the maximum in $\chi''(f)$ at a given temperature). The changing peak position with decreasing temperature reflects the evolution of $\tau(T)$, and our characterization of $\tau(T)$ down to below $T = 1$ K allows us to understand the origins of the two different spin freezing transitions observed in the a.c. susceptibility. As shown in figure 8 and described previously [10,25], $\tau(T)$ displays thermally activated behavior at high temperatures which changes to a much weaker temperature dependence at $T_{cross}$ ~ 13 K. Our data show that the strong temperature dependence then re-emerges below $T_{ice}$, as spin-spin correlations develop. The higher temperature activated relaxation is responsible for the spin freezing observed at $T$ ~ 16 K in the higher frequency a.c. susceptibility data. The crossover to relatively weak temperature dependence results in the absence of freezing at lower frequencies until $\tau(T)$ begins to rise sharply again at the lowest temperatures. This rapid increase of $\tau$ with decreasing temperature is actually faster than would be expected for activated behavior (as shown in the inset to figure 8), which we attribute to the increasingly strong correlations between the spins with decreasing temperature requiring several spins to change orientation in order to follow the a.c. field.

## V. Discussion

With the data presented above, we can contrast the $T < T_{ice}$ spin-freezing with the well-studied transition to a spin glass state. The basic signatures of the spin freezing, i.e. an irreversibility in the magnetization and a frequency dependent maximum in $\chi'(T)$ in the absence of other thermodynamic signatures of a phase transition, are qualitatively consistent with spin-glass freezing seen in both highly disordered systems and in site-ordered geometrically frustrated antiferromagnets [1,2,30]. Upon closer inspection, however, the detailed behavior of the spin-ice freezing is somewhat different from the cooperative freezing in spin glasses, with the most obvious qualitative difference being that the application of a field enhances the freezing temperature. This is in sharp contrast to the behavior of both the disorder-based spin glasses and spin-glass transitions observed in site-ordered geometrically frustrated antiferromagnets [27,31,32].

A more subtle difference between the low temperature spin ice freezing and that in spin glasses is the distribution of spin relaxation times. While this has not been well characterized in the site-ordered geometrically frustrated antiferromagnets, the spin freezing in disorder-based spin glasses is accompanied by an extremely broad distribution of relaxation times [27]. The narrow range of distribution times in $Dy_2Ti_2O_7$, characterized through the width of the peak in $\chi''(f)$, is presumably due to the lack of inhomogeneity in the local environment of individual spins. This lack of disorder combined with the frustration prevents the development of a range of spin-correlation length scales that characterize the low temperature state of spin glasses. Curiously, the peak in $\chi''(f)$ appears to narrow and become noticeably asymmetric at the lowest temperatures, with a much sharper drop on the low frequency side. Similar behavior was observed in $LiHo_{1-x}Y_xF_4$, where it was attributed to the emergence of a gap in the relaxation spectrum [29]. Unfortunately, the frequency range of our apparatus did not allow us to fully characterize the development of this asymmetry at lower temperatures.

The difference between the spin freezing in $Dy_2Ti_2O_7$ and that in spin glasses is perhaps not surprising, since origin of the spin ice state is purely geometrical and does not involve the structural and chemical disorder traditionally associated with glassiness in magnetic materials. The spins in spin ice systems are also highly uniaxial, which makes them rather different in character from those in site-ordered geometrically frustrated antiferromagnets that exhibit spin-freezing, where the spins are typically quite isotropic. These differences suggest the addition of disorder or dilution of the magnetic lattice with magnetic or non-magnetic [8,28,33,34] ions as possible routes for investigating a crossover between spin ice and spin glass behavior. The data suggest that the glassiness observed in $Dy_2Ti_2O_7$ is somehow fundamentally different from that in other magnetic materials, and that different models will be needed to understand the spin freezing. The results also raise the question of whether there exist other types of frustration in site-ordered materials which will manifest glassy behavior with different characteristic behavior.

**ACKNOWLEDGEMENTS**

We gratefully acknowledge support from the Army Research Office PECASE grant DAAD19-01-1-0021. R. J. C. was partially supported by NSF grant DMR-9725979.

**Figure captions**

Figure 1. The temperature dependence of the magnetization measured on warming after zero field cooling and field cooling from $T = 1.5$ K. The increase in the zero field cooled data at the lowest temperatures is due to the non-equilibrium effects associated with raising the magnetic field at low temperatures.

Figure 2. Magnetization as a function of field at 250 mK, 400 mK, and 800 mK showing hysteresis below $T \sim 650$ mK but completely reversible behavior at higher temperatures. The black curve is the initial sweep up in field after zero-field cooling, followed by the red and green curves respectively.

Figure 3. IRM and TRM data at 250 mK showing the presence of irreversibility and the formation of a metastable state at low temperatures. Notice that they merge in fields above 5 kOe where the bifurcation between FC and ZFC magnetization also disappears.

Figure 4. The ac susceptibility of $Dy_2Ti_2O_7$ as a function of temperature in zero magnetic field. a) The real part of the susceptibility ($\chi'$) showing frequency dependent local maximum at $\sim 4$ K and $\sim 16$ K. b) The imaginary part of the susceptibility ($\chi''$) showing rises at temperatures corresponding to the drops seem in $\chi'(T)$.

Figure 5. The temperature dependence of the real and imaginary parts of the ac susceptibility at low temperatures in zero magnetic field.

Figure 6. The frequency of the spin freezing temperature below $T_{ice}$. Note that the data do not follow Arrhenius behavior and that the application of a magnetic field increases the freezing temperature. The open symbols represent zero field data taken in the dilution refrigerator while the filled symbols represent the higher temperature data taken on the PPMS cryostat.

Figure 7. The imaginary part of the ac susceptibility as a function of frequency at low temperatures in zero applied field. The prominent single peak in the data suggests that they are well-described by a single characteristic relaxation time.

Figure 8. The temperature dependence of the characteristic spin relaxation time. The open symbols represent zero field data taken in the dilution refrigerator while the filled symbols represent the higher temperature data taken on the PPMS cryostat. The inset shows low temperature ($T < 4$ K) data plotted as a function of $1/T$ which shows the non-Arrhenius behavior at low temperatures. Note that $\tau(T)$ is increasing at a rate which is faster than exponential in $1/T$, i.e. the increase is faster than what would be expected for simple thermal activation.

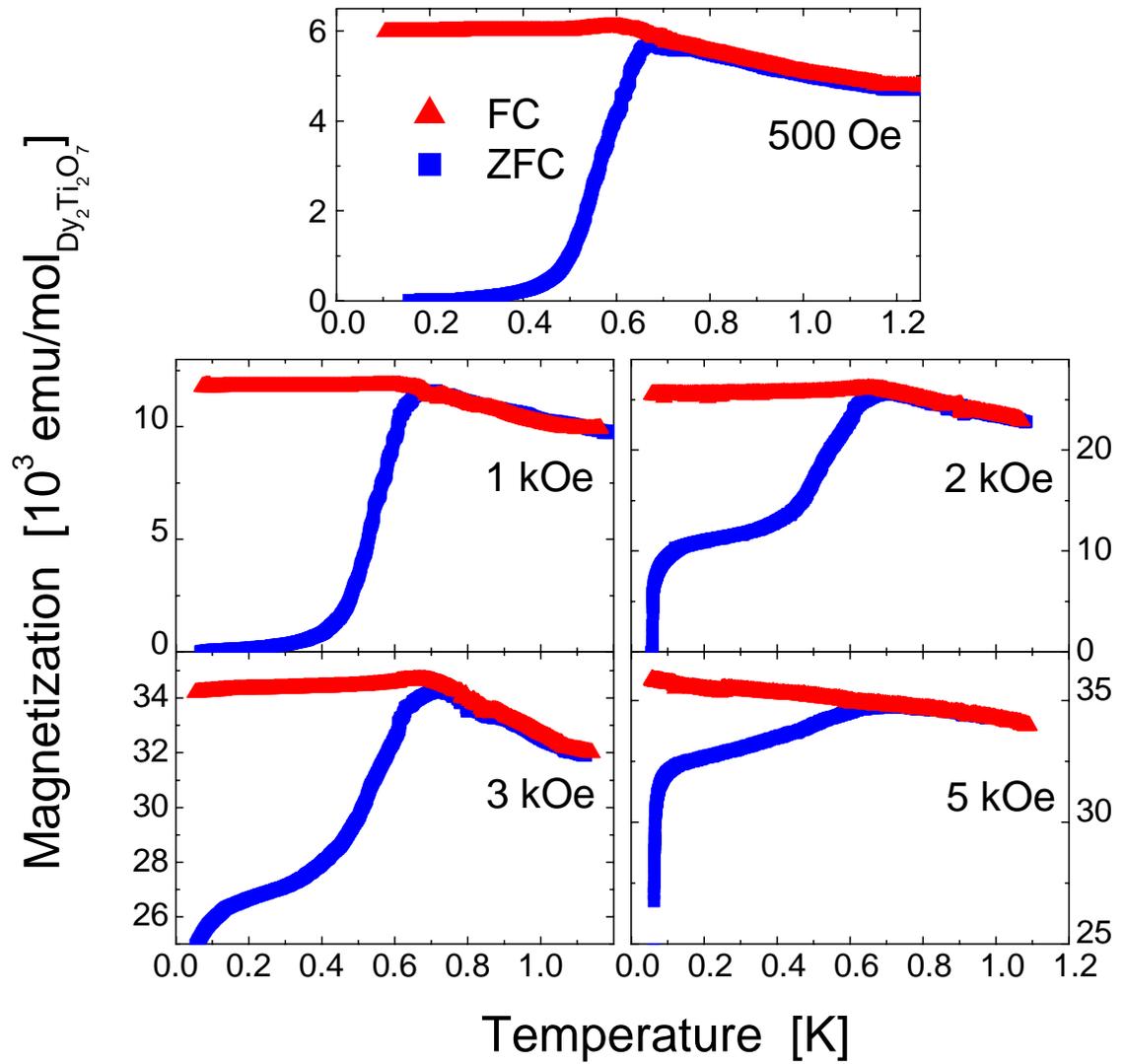

figure 2 Snyder *et al.*

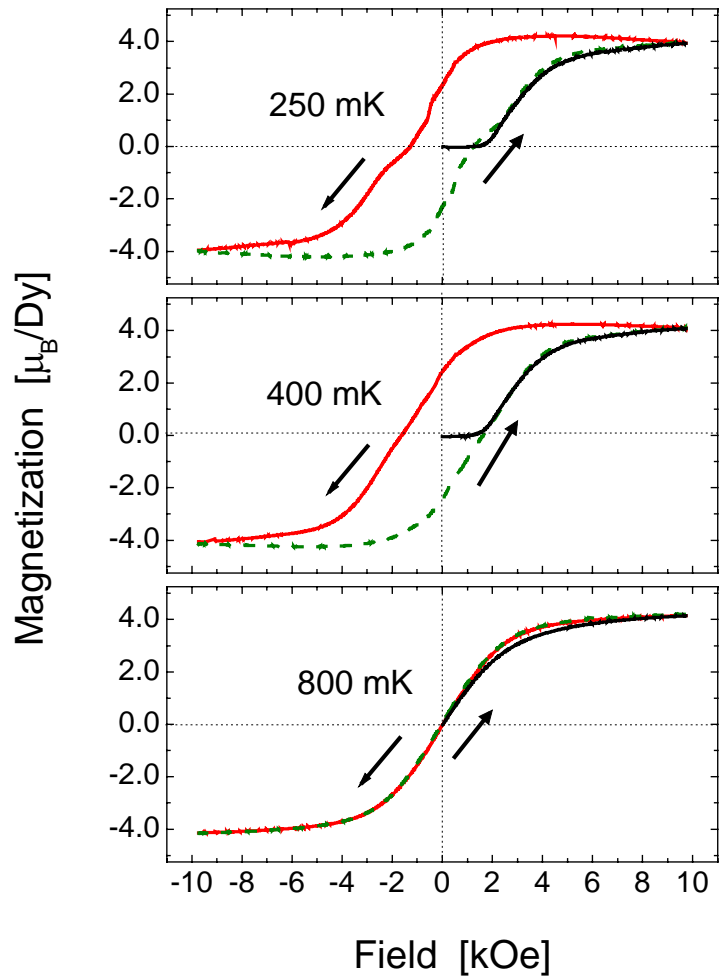

Figure 3 Snyder *et al*.

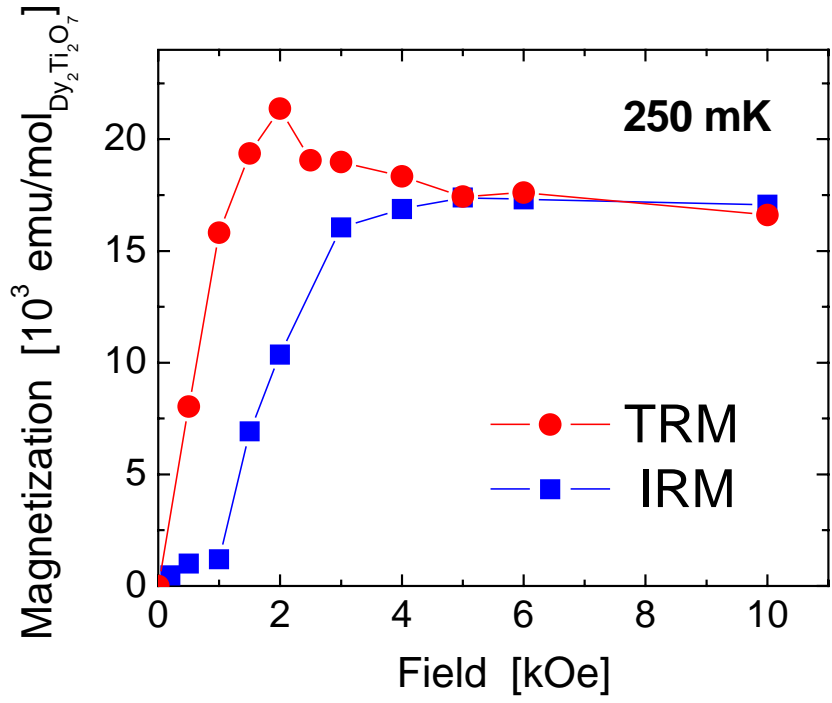

Figure 4 Snyder *et al.*

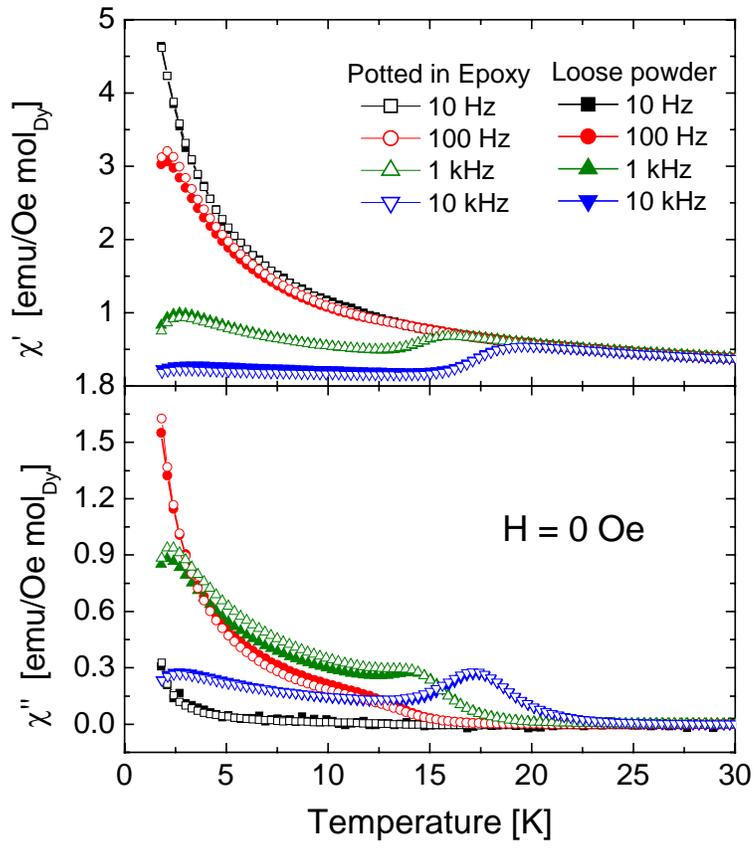

Figure 5 Snyder *et al.*

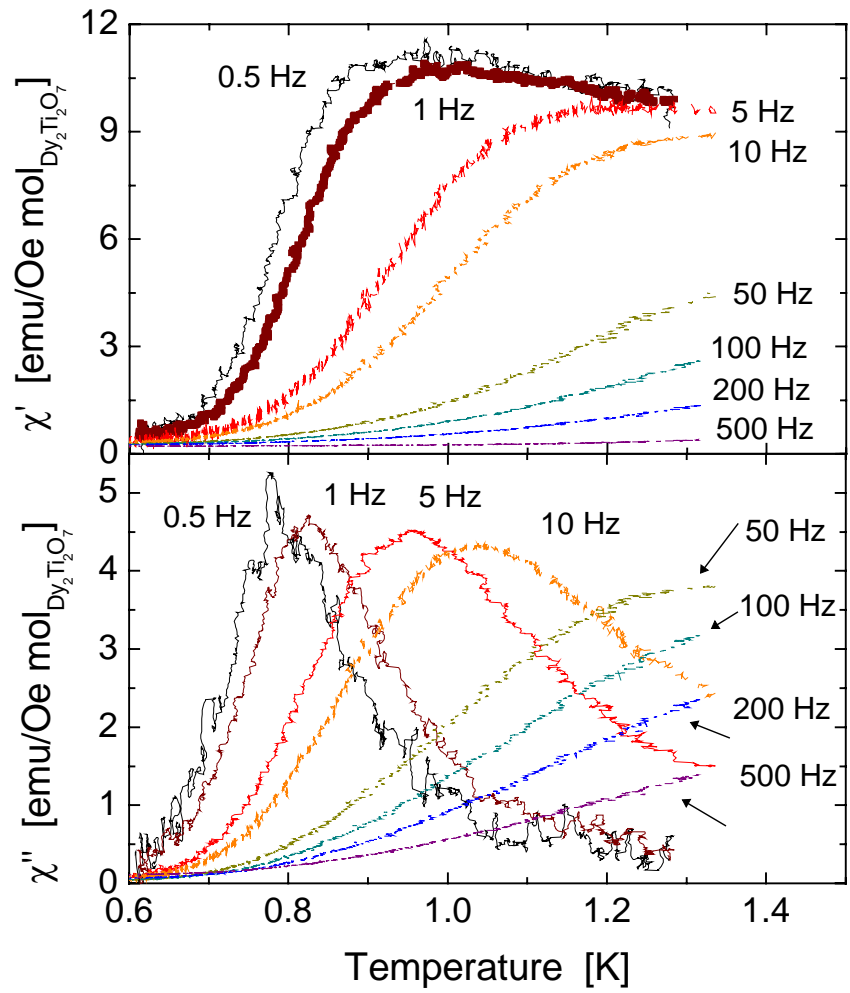

Figure 6 Snyder *et al.*

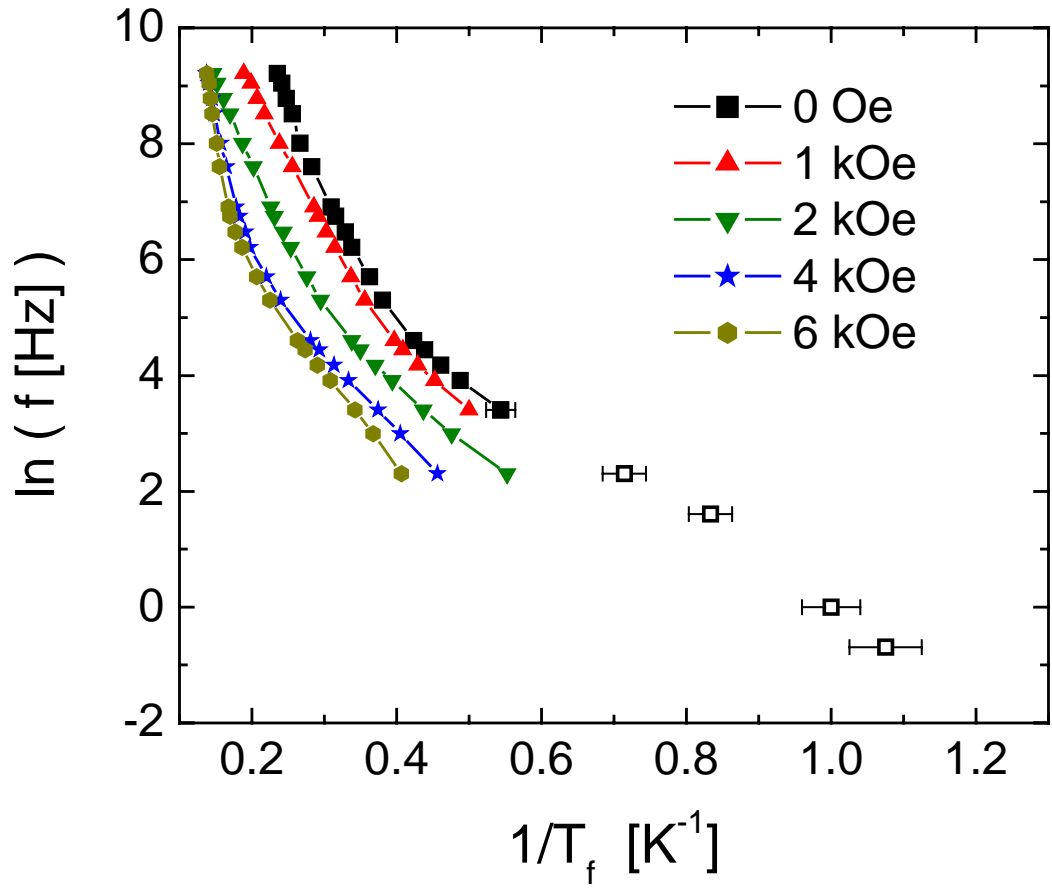

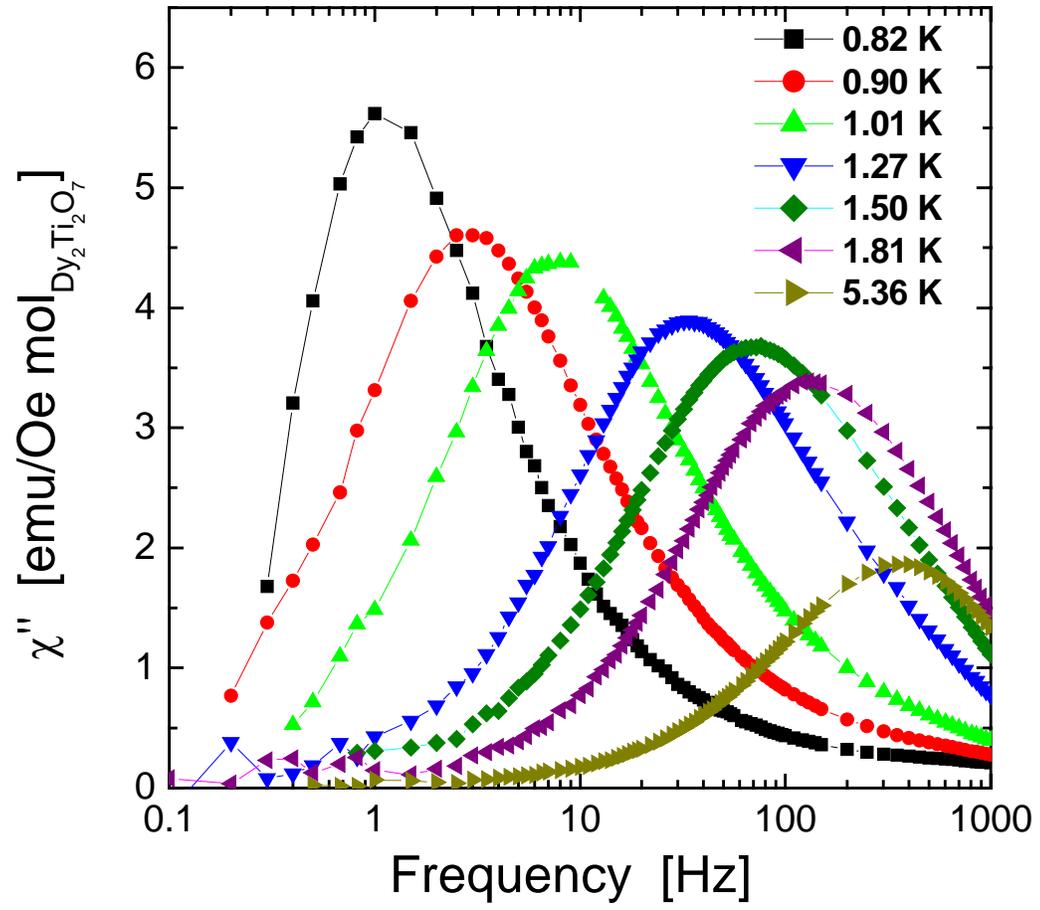

Figure 8   Snyder *et al.*

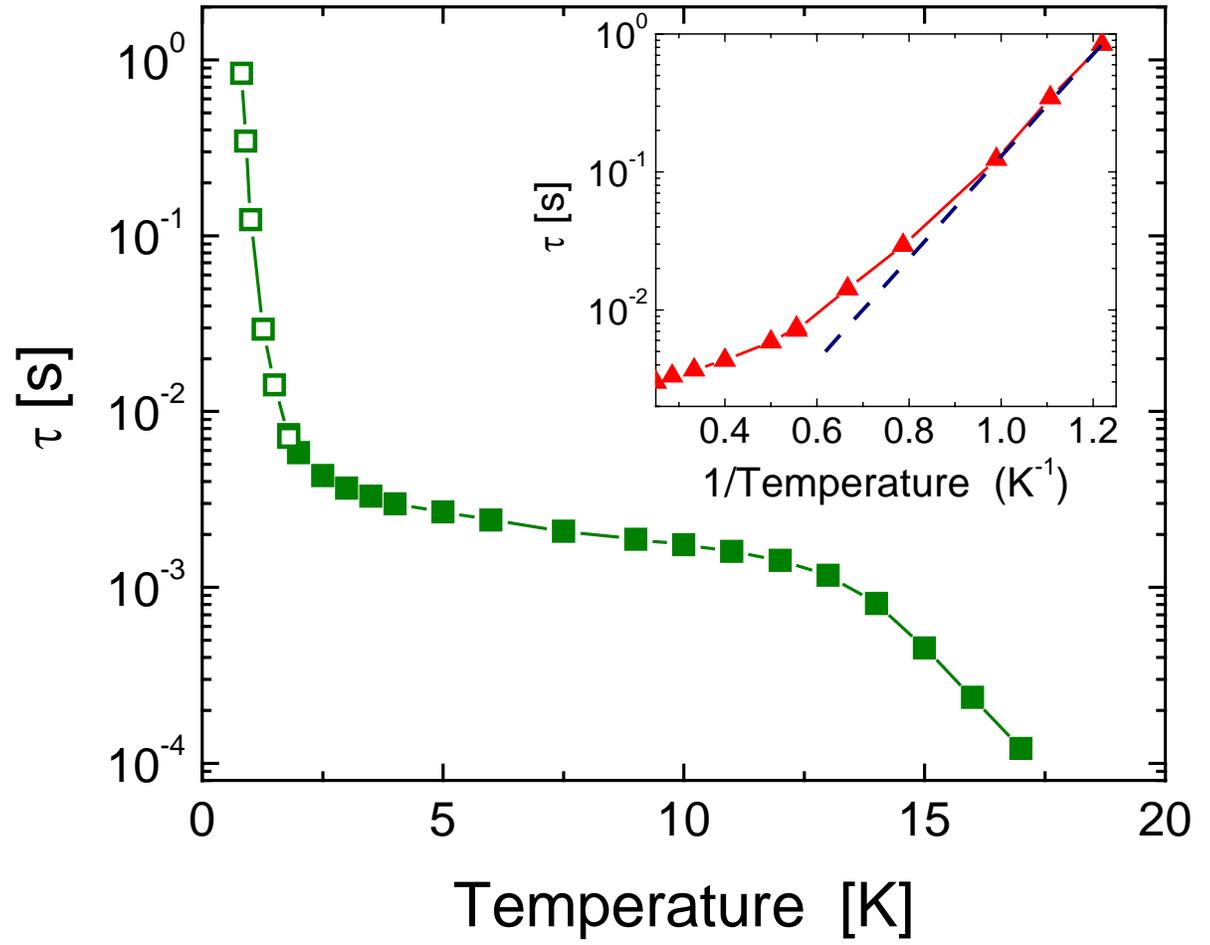